# Intrinsic dielectric properties of magnetodielectric $La_2CoMnO_6$

R. X. Silva[1,2], R. L. Moreira[3], R. M. Almeida[3], R. Paniago[3] and C. W. A. Paschoal[2,*]

[1] Departamento de Física, Universidade Federal do Maranhão, Campos VII, 65400-000, Codó - MA, Brazil

[2] Departamento de Física, Universidade Federal do Maranhão, Campus do Bacanga, 65085-580, São Luís - MA, Brazil

[3] Departamento de Física, Universidade Federal de Minas Gerais, ICEx, 31270-901 Belo Horizonte - MG, Brazil

[*] Corresponding author. Tel: +55 98 3301 8291

E-mail address: paschoal.william@gmail.com (C. W. A Paschoal)

**Abstract**

Manganites with double perovskite structure are being attractive because of their interesting magnetoelectric and dielectric responses. Particularly, colossal dielectric constant (CDC) behavior has been observed in $La_2CoMnO_6$ (LCMO), at radio frequencies and room temperature. In this paper, we employed infrared-reflectivity spectroscopy onto a LCMO ceramic obtained through a modified Pechini's method to determine the phonon contribution to the intrinsic dielectric response of the system and to investigate the CDC origin. The analysis of the main polar modes and of the obtained phonon parameters show that CDC effect of LCMO is of pure extrinsic origin. In addition, we have estimated the dielectric constant and quality factor of the material in microwave region, $\varepsilon'_s \sim 16$ and $Q_u \times f \sim 124$ THz, which shows that LCMO is appropriate for applications into microwave devices and circuitry.

# 1. Introduction

In the past fifteen years, double perovskites with chemical formula $RE_2MeMnO_6$ (where RE = rare earth; and Me = Ni or Co) have attracted a lot of attention because they can present very interesting physical properties, such as multiferroicity[1,2], and a variety of electrical/magnetic couplings, with potential applications in new technologies[2,3]. In particular, $La_2CoMnO_6$ (LCMO) and $La_2NiMnO_6$ (LNMO) has a distinguished position among other members of the series, mainly because of their relative higher Curie temperature ($T_c$ ~ 230K and ~ 280K, respectively), and strong electrical-magnetic coupled phenomena like magnetoelectric (ME)[4], magnetodielectric (MD)[5–8], magnetoresistive (MR)[9] and colossal magnetoresistive (MCR)[10] effects.

Structural investigations by neutron diffraction has shown that these compounds crystallize into a monoclinic structure with charge ordering of *rock salt* type ($A_2BB'O_6$), belonging to the $P2_1/n$, (#14 or $C_{2h}^5$)[11,12] space group. Within this structure, the cationic ordering state of the B e B' ions, occupying respectively the 2c e 2d Wyckoff sites, has a strong influence on the vibrational, electrical and magnetic properties of the material. In particular, the ordering of the magnetic ions (Mn, Co or Ni) in B/B' sites affects the spin-phonon coupling[13,14], the critical magnetic temperatures[15], the electrical conductivity[12,16] and the dielectric constant[17,18]. Concerning the dielectric constant of LCMO there are some discrepancies on the results presented in the literature: Yáñez-Vilar *et al.* [18] reported that the radio-frequency (<$10^4$ Hz) dielectric constant of their ceramic samples, at room temperature, increased by two orders of magnitude from the ordered to disordered material, attaining values as high as $10^5$. By increasing the frequency ($10^6$ Hz) and lowering the temperature (100 K), the dielectric constant decreases to $\varepsilon'$ ~ 15, for the ordered and $\varepsilon'$ ~ 30, for the disordered samples. Barón-González *et al.* [12], on the same external conditions ($10^6$ Hz and

100 K), obtained $\varepsilon'$ ~67 and $\varepsilon'$ ~28, for their ordered and disordered LCMO ceramic samples, respectively.

Complex impedance analyses performed by Lin and Chen[19] demonstrated clearly the colossal dielectric constant (CDC) behavior of LCMO ceramics. Indeed, the dielectric constant relaxes from about $10^5$ at low frequencies ( ~ 10 Hz), to values *ca.* $10^2$ at high frequencies (~$10^7$ Hz). The observed dispersion (Debye-like relaxation) is quite similar to that observed in $CaCu_3Ti_4O_{12}$[20,21], where extrinsic defects from intentional or non-intentional doping have been considered as responsible for the intriguing CDC response[22]. On the other hand, the relaxed high frequency values are close to the intrinsic dielectric constant of the pure material[22]. In addition, Lin and Chen[19] revealed a relaxor-like behavior of their samples, which presented a dependence with $Co^{2+}/Mn^{4+}$ cationic ordering, as does the CDC effect. Later on, Venimadhav *et al.*[23] obtained values of *ca.* 10 for the dielectric constant at low temperatures (below 20 K) for LCMO nanoparticles with typical dimensions of *ca.* 28 nm. By analyzing deviations on the impedance arcs of their samples, they concluded that intrinsic and extrinsic features could contribute to the MD effect in LCMO. Indeed, this effect can have different origins, depending on the electronic and phononic structures of the material.

Infrared spectroscopy is an ideal tool for solving controversies concerning intrinsic or extrinsic nature of the dielectric response of an insulator (and/or high-gap semiconductor) material, because for very low charged defects concentration, the infrared dielectric response ($10^{11}$-$10^{14}$ Hz) is completely determined from the phononic and electronic polarizations, i.e., is purely of intrinsic origin. Therefore, this technique would be very helpful for investigating the dielectric behavior of LCMO and discuss its CDC effect.

In addition, complex double perovskites with 1:1 B-site charge ordering are usually developed and investigated for microwave (MW) applications as dielectric resonators and filters, due to appropriate dielectric constant and low dielectric losses in the MW region, which are improved with charging ordering[24,25]. Since these materials usually belong to centro-symmetric groups, the investigation of their optical polar phonons by infrared spectroscopy for determining their intrinsic dielectric responses at MW region is mandatory, once these modes are mutually exclusive to the Raman ones through selection rules. At the best of our knowledge, recent spectroscopic studies on phonon behavior of materials such as LCMO and LNMO considered only vibrational analyses of Raman-active (non-polar) modes[13,26].

In order to determine the polar phonon features and their contributions to the intrinsic dielectric response of LCMO-type materials, we have investigated the infrared reflectivity spectra of a LCMO ceramic obtained by a modified Pechini's method. The obtained results allowed us to elucidate which vibrational modes give an effective contribution for the MW dielectric response of the material, and, therefore, to discuss the origin of the CDC effect.

## 2. Experimental procedures

$La_2CoMnO_6$ polycrystalline sample was obtained from a modified polymeric precursor (MPP) route based on Pechini's method[27], using cobalt acetate tetrahydrate ($C_4H_6CoO_4 \cdot 4H_2O$, Sigma Aldrich Co.), manganese nitrate hydrate ($MnN_2O_6 \cdot xH_2O$, Sigma Aldrich) and high purity lanthanum oxide ($La_2O_3$, Sigma, Aldrich) as metal sources to obtain the precursors. The precursors were mixed to obtain the $La_2CoMnO_6$ perovskite. The obtained resin was annealed at 400°C for 2h resulting in a black porous powder that was grinded in an agate mortar before calcination for 16 hours at 1000°C. After that, the material was regrind, pressed and annealed again at 1000 °C, for the same period. The crystal structure of the sample was probed by X-ray powder diffraction (XRPD) using a Bruker D8 Advance with

Cu-Kα radiation (40 kV, 40 mA), over a range from 10° to 100° (0.02°/step with 0.3s/step). The XRPD pattern was compared with data from ICSD (Inorganic Crystal Structure Database, FIZ Karlsruhe and NIST) International diffraction database (ICSD# 98249) [28]. The structure was refined using the GSAS code[29,30].

X-ray photoelectron spectroscopic (XPS) measurements were performed in a VG ESCALAB 220i-XL system, using Al-K$_α$ radiation (1486.6 eV) and base pressure of 1 x 10$^{-10}$ mbar. Survey XPS spectra were collected with pass energy of 50 eV, while Co 2p and Mn 2p spectra with 20 pass energy.

Infrared reflectivity spectra were collected in a Fourier-transform spectrometer (FTIR) Bomem DA8-02 equipped with a fixed-angle specular reflectance accessory (external incidence angle of 11.5 °). In the mid-infrared region (500 – 4000 cm$^{-1}$) a SiC Globar lamp as infrared source, a Ge-coated KBr beamsplitter and LN$_2$-cooled HgCdTe detector were used. In the far-infrared range (50 – 600 cm$^{-1}$), a mercury-arc lamp, a 6 mm coated Mylar hypersplitter®, and LHe-cooled Si bolometer were employed. For obtaining the reference spectra, one region of the ceramic face was covered with a thin gold coating, simulating a ''rough'' mirror. This method helped us to improve the reflectivity spectra, since the mirror surface mimics the sample one, compensating the losses due to diffuse reflection at the rough sample surface.

## 3. Results and discussions

The room-temperature XRPD pattern obtained for the LCMO ceramic sample is shown in Figure 1. LCMO has a monoclinic symmetry, belonging to the P2$_1$/n (#14 ou C$_{2h}^5$) space group, compatible with a 1:1 charge ordering of Mn$^{4+}$/Co$^{2+}$ ions on the B-sites. Refinement parameters are in excellent agreement with obtained with SPuDS[31] program, and with reported results on LCMO sample obtained by a modified route of nitrates decomposition and thermally treated at 1100°C for 16h[32]. Within the detection limit of the diffraction pattern, the sample was impurity free, with no $La_2O_3$ traces, as usual in rare earth manganites with double perovskite structure[33,34]. The obtained lattice parameters were $a$ = 5.5255(1) Å, $b$ = 5.4847(1) Å, $c$ = 7.7771(2) and $\beta$ = 89.926°, in good agreement with the literature[11]. Details on structural analysis and refinement parameters are given in the Supplemental Material[35].

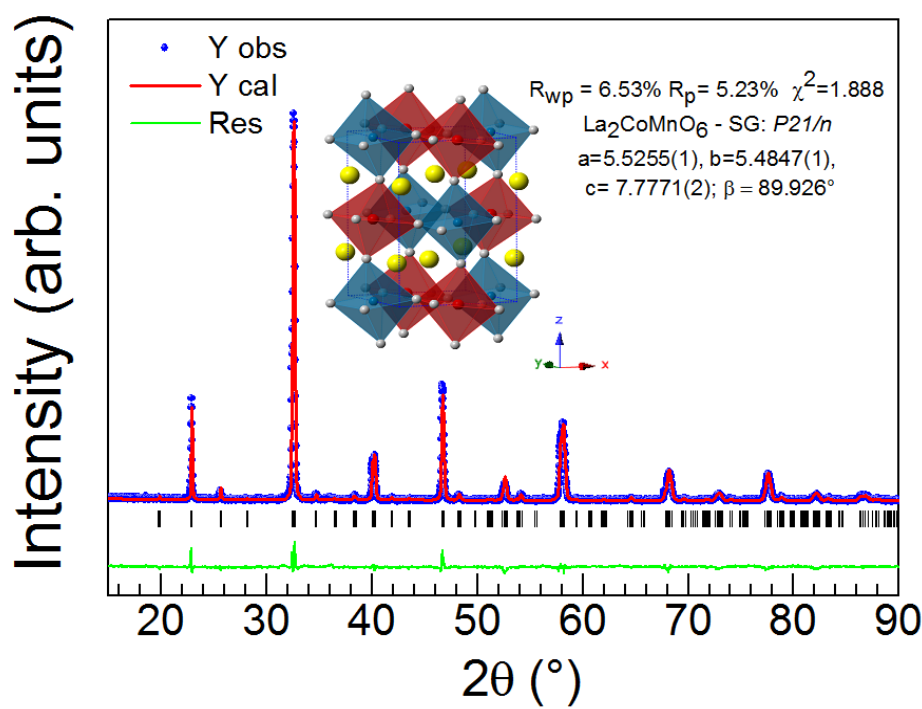

Figure 1 – XRPD pattern of LCMO ceramic sintered at 1000 °C for 16 h (blue dots). The red solid line is the fitting using the Rietveld method and the green line is the residual between the experimental and calculated patterns. The inset shows the LCMO structure.

In order to investigate the electronic structure and oxidation states (Ox.S) of Co and Mn ions in LCMO we performed XPS measurements. The spectra collected in the Co 2p and Mn 2p binding energy regions are shown in Figures 2 (a) and (b), respectively. The main peak in the spectrum of Co $2p_{3/2}$ is centered around 780,2 eV, while in CoO, in which cobalt has a

2+ oxidation state, the corresponding peak appears at 780,4 eV[36]. The satellite peak observed around 787,1 eV is also characteristic of Co $2p_{3/2}$, being present in XPS spectra of CoO[37], likewise. Figure 2 (b) shows that the core-level Mn $2p_{3/2}$ peak is centered at 642,4 eV; the corresponding peak in $MnO_2$ is observed at 642,2 eV[36]. The XPS results altogether demonstrate that the oxidation states of Co and Mn are essentially $Co^{2+}/Mn^{4+}$. The charge difference for these ions is coherent with charge ordering on B-site and compatible with the monoclinic structure identified by the X-ray measurements.

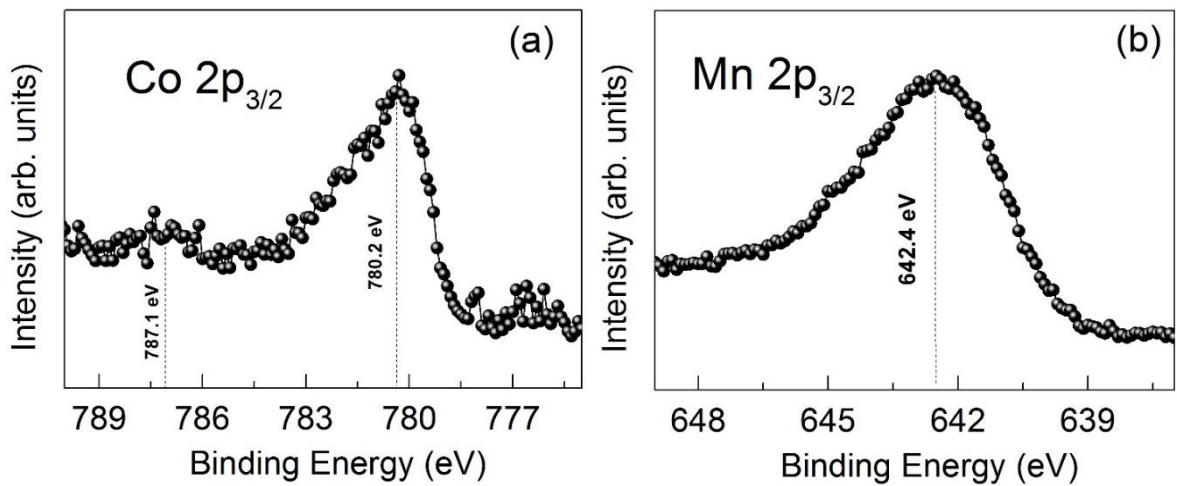

Figure 2 – XPS spectra of LCMO sample: (a) in Co 2p and (b) Mn 2p binding energy regions.

By considering the symmetries and Wyckoff site occupations in LCMO structure within the $P2_1/n$, (#14 or $C_{2h}^5$)[11,12] space group, with two chemical formula by unit cell ($Z = 2$), group theoretical tools help us to predict, in terms of irreducible representation of the $C_{2h}$ factor group, 24 Raman-active modes (12$A_g$ ⊕ 12$B_g$), and 33 polar infrared-active (IR) modes, (17$A_u$ ⊕ 16$B_u$), as shown in Table I. As it is well known, for the ideal cubic perovskite of general formula $ABO_3$, belonging to the $Pm\bar{3}m$ ($O_h^1$ or #221) space group, only three IR-active modes are predicted, namely: an external mode due to the relative movement of the A ions to the $BO_6$ sub-lattices, with frequencies in the 100 e 170 $cm^{-1}$ range; the O-B-O bending mode with frequency in the 200-500 $cm^{-1}$ range; and one stretching mode above 500

cm$^{-1}$, due to octahedra vibrations[38,39]. Concerning the complex perovskites of general formula $A_2BB'O_6$ with no *tilts* (Glazer notation a$^0$ a$^0$ a$^0$), the $Fm\bar{3}m$ ($O_h^5$ or #225) space group is usually observed[40]. In this structure, group theory predicts 4 Raman ($A_g \oplus E_g \oplus 2F_{2g}$), and 4 distinct IR-active modes ($4F_{1u}$). However, elastic distortions coming from the B-site occupation, along with *tilts* for accommodating the rare-earth ion and A-O distance optimization lead to crystal symmetry lowering, with consequent increasing of the number of first-order Raman and infrared-active normal modes. It is worth mentioning that, although the Wyckoff sites occupied by Mn and Co ions in monoclinic LCMO were different, they have the same local symmetry ($C_i$), which do not contribute to the Raman spectra, but participate uniquely of the IR-active polar modes (see Table I).

Table I – Normal vibrational modes of LCMO at the Brillouin zone center (Γ-point), within the $\boldsymbol{P2_1/n}$, (#14 or $\boldsymbol{C_{2h}^5}$) monoclinic space group, with Z=2.

| Atom | Ox. State | Wyckoff position | Occup. fraction | Site symmetry | Irreducible representations |
|---|---|---|---|---|---|
| La | 3+ | 4e | 1 | $C_1$ | $3A_g \oplus 3A_u \oplus 3B_g \oplus 3B_u$ |
| Co | 2+ | 2c | 1 | $C_i$ | $3A_u \oplus 3B_u$ |
| Mn | 4+ | 2d | 1 | $C_i$ | $3A_u \oplus 3B_u$ |
| O$_{(1)}$ | 2- | 4e | 1 | $C_1$ | $3A_g \oplus 3A_u \oplus 3B_g \oplus 3B_u$ |
| O$_{(2)}$ | 2- | 4e | 1 | $C_1$ | $3A_g \oplus 3A_u \oplus 3B_g \oplus 3B_u$ |
| O$_{(3)}$ | 2- | 4e | 1 | $C_1$ | $3A_g \oplus 3A_u \oplus 3B_g \oplus 3B_u$ |
| | | | | | $\Gamma_{IR} = 17A_u \oplus 16B_u$ |
| | | | | | $\Gamma_{Raman} = 12A_g \oplus 12B_g$ |
| | | | | | $\Gamma_{acoustic} = A_u \oplus 2B_u$ |

Although group theory foresees 33 polar modes for the monoclinic LCMO, the FTIR reflectivity spectra of the material, presented in Figure 3 (as open black circles), shows rather a group of 4 broad bands, which could be resolved with 13 modes, as discussed later. Visually, the spectrum resembles those for cubic (undistorted) double pervskites, with their observed 4 bands. With increasing elastic distortion and tolerance factor reduction, the splitting of the degenerate and activation of additional first order modes become more effective and it is possible to visualize new bands and shoulders in the lower symmetry system. However, the relatively small monoclinic distortion along with the absence of preferencial orientations in ceramic samples (that are nearly perfect random polycrystals) forbid that the $A_u$ and $B_u$ modes could be resolved by polarisation. Thus, as it is a common case for monoclinic 1:1 ordered double perovskites, a maximum of 17 bands (corresponding to unresolved $A_u$ and $B_u$ modes) could be expected for our system [25,40–42]. The actual spectrum may contain a lower number of oscillators, since the intensities of some modes could be too weak and their positions too close to strongest modes, so that the weaker modes become difficult to be discerned[24,43].

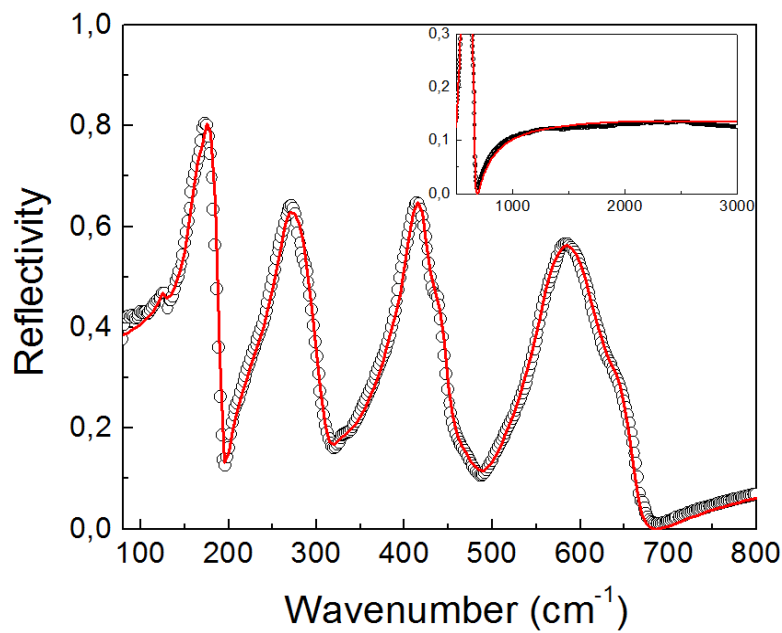

Figure 3 – Infrared reflectivity spectrum of the LCMO ceramic, at room temperature. The black open circles represent experimental data and red line the fitted spectrum from the four-parameter semi-quantum model. The inset shows the corresponding spectra in the mid-infrared region.

The FTIR spectrum of our LCMO ceramic samples was analyzed based on the four-parameter semi-quantum model proposed by Gervais and Piriou[44]. According to this model, the complex dielectric function $\varepsilon(\omega)$ is expressed in therms of the IR-active phonons as follows

$$\varepsilon(\omega) = \varepsilon_\infty \prod_j \frac{\left(\omega_{j,LO}^2 - \omega^2 + i\omega\gamma_{j,LO}\right)}{\left(\omega_{j,TO}^2 - \omega^2 + i\omega\gamma_{j,TO}\right)} , \qquad (1)$$

where $\omega_{j,TO}$ e $\omega_{j,LO}$ correspond to the frequencies of the optical transversal (TO) and longitudinal (LO) branches of the $j$-th mode, respectively, $\gamma_{j,TO}$ e $\gamma_{j,LO}$ are the corresponding damping constants, and $\varepsilon_\infty$ is the high frequency dielectric constant due to the electronic polarization. The infrared reflectivity spectrum is fitted with Equation (1) together with the Fresnel formula, which under quasi-normal incidence can be written as

$$R(\omega) = \left|\frac{\sqrt{\varepsilon(\omega)}-1}{\sqrt{\varepsilon(\omega)}+1}\right|^2 . \qquad (2)$$

The red line in Figure 3 shows the fitted curve for the reflectivity data of LCMO (black open circles), in the spectral region 80–3000 cm$^{-1}$, obtained within the model outlined above. The initial fitting parameters were obtained from Kramers-Kronig analysis of the experimental data. The complex dielectric function, $\varepsilon(\omega) = \varepsilon'(\omega) - i\varepsilon''(\omega)$ , has been calculated from the fitting of the reflectivity spectrum. The optical functions $\varepsilon'$, $\varepsilon''$ and Im(1/$\varepsilon$) obtained from this fit are shown in Figure 4. The imaginary functions $\varepsilon''$ and Im(1/$\varepsilon$) provide the positions and damping constants of TO and LO branches, respectively.

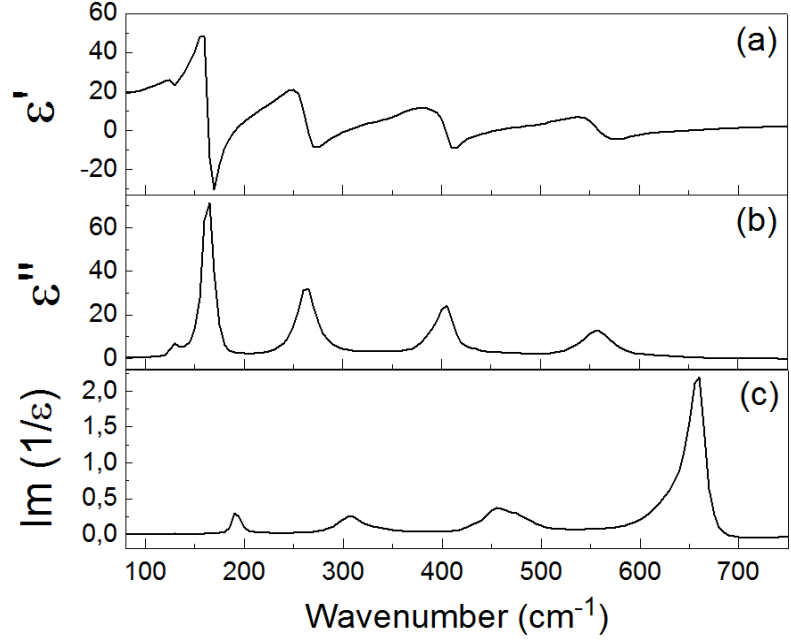

Figure 4 – (a) Real, and (b) imaginary parts of the dielectric functions of a LCMO ceramic in the far infrared region; (c) imaginary part of the reciprocal dielectric function.

Table II presents the complete fitting parameters set for LCMO. The table gives the individual phonon wavenumbers, damping constants, dielectric strengths ($\Delta\varepsilon_j$) and losses (tan$^{-1}$ $\delta_j$), as well as the extrapolated *static* dielectric constant ($\varepsilon_s$) and quality factor ($Q_u$), at 10 GHz. In terms of the optical phonon parameters, the dielectric strengths can be obtained from

$$\Delta\varepsilon_{j,TO} = \frac{\varepsilon_\infty}{\omega_{jTO}^2} \times \frac{\Pi_k(\omega_{k,LO}^2 - \omega_{j,TO}^2)}{\Pi_{k \neq j}(\omega_{k,TO}^2 - \omega_{j,TO}^2)} \qquad . \qquad (3)$$

Table II – Dispersion parameters from the best numerical fit of the FTIR spectrum of LCMO ceramic.

| Modes | $\omega_{j,TO}$ (cm$^{-1}$) | $\Upsilon_{j,TO}$ (cm$^{-1}$) | $\omega_{j,LO}$ (cm$^{-1}$) | $\Upsilon_{j,LO}$ (cm$^{-1}$) | $\Delta\varepsilon_j$ | $10^8 \tan\delta_j/\omega$ |
|---|---|---|---|---|---|---|
| 1 | 127.6 | 8 | 128.4 | 9 | 0.30 | 932 |
| 2 | 163.4 | 12 | 172.6 | 18 | 5.03 | 14500 |
| 3 | 173.0 | 12 | 191.7 | 9 | 0.14 | 368 |
| 4 | 262.9 | 23 | 283.9 | 34 | 2.87 | 6015 |
| 5 | 284.4 | 35 | 307.2 | 30 | 0.04 | 99 |
| 6 | 329.0 | 30 | 331.0 | 31 | 0.05 | 87 |
| 7 | 385.6 | 29 | 386.5 | 33 | 0.08 | 92 |
| 8 | 406.4 | 21 | 427.5 | 24 | 1.13 | 897 |

| | | | | | |
|---|---|---|---|---|---|
| 9 | 433.8 | 27 | 449.9 | 28 | 0.16 | 144 |
| 10 | 470.2 | 66 | 482.4 | 45 | 0.17 | 317 |
| 11 | 558.2 | 39 | 598.7 | 57 | 0.94 | 746 |
| 12 | 599.0 | 59 | 621.4 | 55 | $< 10^{-2}$ | 1 |
| 13 | 634.0 | 66 | 660.2 | 19 | 0.05 | 54 |
| $\varepsilon_\infty = 4.90$   $\Sigma \Delta\varepsilon_j = 10.94$ | | | $\Sigma \tan \delta_j / \omega = 24252 \times 10^{-8}$ | | |
| $\varepsilon_s = \varepsilon_\infty + \Sigma \Delta\varepsilon_j = 15.84$ | | | $Q_u \times f = 124$ THz | | |

The infrared spectrum of LCMO can be divided up into four groups of bands, as described in the sequence. The first group (I) has three modes of low wavenumber, into the region 100–180 cm$^{-1}$. These vibrations can be attributed to the motion of La$^{3+}$ cations related to (Mn/Co)O$_6$ octahedra. Discarding the electronic polarization, these modes contribute with about 50% of the *static* (infrared) dielectric constant (mode #2 has the strongest dielectric strength, with $\Delta\varepsilon_2 = 5.03$). Therefore, these low frequency atomic vibrations generate the highest dipole moment of the polar phonon spectrum of LCMO. Two intermediate frequency groups, in the (II) 250–330 cm$^{-1}$ range, with three vibrational modes, and (III) 370–480 cm$^{-1}$ region, with four modes, contribute respectively for the *static* dielectric constant with 27% and 14%. The strongest modes of these groups are those numbered as #4 and #8. These intermediate frequency bands are attributed to bending modes of O–(Mn/Co)–O entities. The latest group of atomic vibrations, (IV) with wavenumbers ω>500 cm$^{-1}$, is associated to (Mn/Co)O$_6$ octahedral stretching modes. This spectral region presents the lowest contribution for the *static* dielectric constant, of only about 9%, which comes essentially from mode #11.

In order to investigate in more detail the contribution of the polar phonons to the dielectric response of the system, we show in Figure 5 the representation of the eigenvectors of the polar active modes whose eigenvalues match well with the wavenumbers of the highest dielectric strengths modes, i.e., modes #2, #4, #8 and #11. Since the reflectivity spectra of LCMO ceramic were depolarized, it was impossible to discern $A_u$ from $B_u$ modes. Thus, pairs of likely modes with both symmetries are presented in Figure 5, for each of the modes #2, #4, #8 and #11. The eigenvectors were calculated from lattice dynamics by the Wilson's GF-matrix method[45,46]. The modeling was optimized by applying a least square route to Raman spectroscopic data[14,26], with discrepancy lower than 5%. The method does not include long range electrostatic interactions, so the calculated results cannot reproduce TO-LO splitting experimentally observed, and only the frequencies of the TO branches are obtained. Details on lattice dynamics calculations are presented in the Supplemental Material[35].

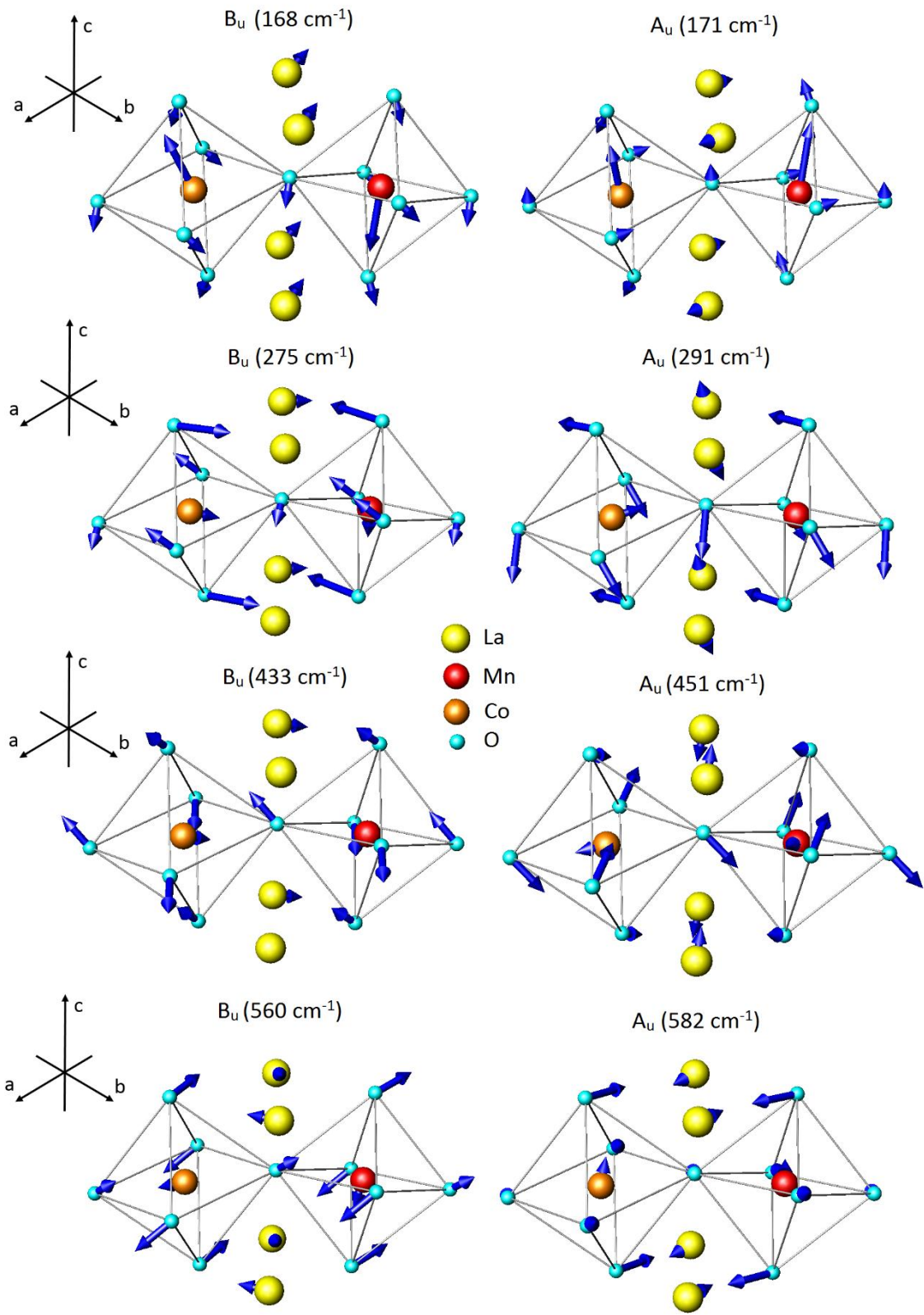

Figure 5 – Sketch of the four strongest IR-active vibrational modes of LCMO (highest $\Delta\varepsilon$), as obtained from lattice dynamic calculations (modes #2, #4, #8 and #11 of Table II). The yellow, orange, red and blue circles represent lanthanum, cobalt, manganese and oxygen ions, respectively.

One can notice, from Figure 5, that the atomic displacements of Mn and Co ions in the low-frequency modes are relatively large. The motion of these ions, in opposite directions relative to the oxygen ions, leads to high dipole moments that, in general, are responsible for the high oscillator (and dielectric) strengths. The high $\Delta\varepsilon$ value for mode #2 denotes the importance of B-site charge ordering for the *static* (infrared) dielectric constant, which must assume different values, depending on the ordering of these ions. On the other hand, complex motions of the oxygen ions, which occupy low symmetry Wyckoff sites with no restrictions concerning their movement directions, dominate the higher frequency modes. In such cases, the resulting dipolar moments are small, and so are the corresponding $\Delta\varepsilon_j$ values.

Most of the polar phonons present relatively small values for $\Delta\varepsilon_j$, and small contributions for the intrinsic (MW) dielectric losses $tan\ \delta_j$, which are obtained from

$$tan\ \delta_j = \omega \frac{\Delta\varepsilon_j \gamma_{j,TO}/\omega_{j,TO}^2}{\varepsilon_\infty + \Sigma_j \Delta\varepsilon_j}. \tag{4}$$

The calculated dielectric losses for the individual phonons from Eq. (4) are also presented in Table II. From these values, the estimated unloaded quality factor in the microwave region (at $f = 10$ GHz or 0.3333 cm$^{-1}$) for our ordered LCMO ceramic is

$$Q_u \times f\ =\ \varepsilon'/\varepsilon'' \times f\ =\ \tan^{-1}\delta \times f \sim 124\ \text{THz}\quad. \tag{5}$$

This result shows that this material has relatively a high MW quality factor, quite appropriate for applications in MW circuitry.

The *static* infrared dielectric constant, $\varepsilon_s$, that is the extrapolated value of the $\varepsilon'$ at the MW region (usually 10 GHz), can be considered as frequency independent and is given by

$$\varepsilon_s =\ \varepsilon'(\omega \rightarrow 0) = \varepsilon_\infty + \Sigma\Delta\varepsilon_j \tag{6}$$

once $\omega_{j,LO} \gg \omega$. As discussed earlier, $\varepsilon_s$ corresponds to the intrinsic dielectric constant of the bulk material. Therefore, four our ordered LCMO material, we have obtained an intrinsic dielectric constant $\varepsilon'_{intrisic} \sim 16$, which is well compatible with the centro-symmetric nature of its structure. The obtained *static* infrared constant value is quite close to values obtained by Yáñez-Vilar *et al.*[18], who found $\varepsilon'$ ~ 15 at radio-frequency (> $10^4$ Hz) low temperatures (100K) for ordered LCMO, and by other authors, who obtained $\varepsilon'$ around 10 at high frequencies[19] and low temperatures[23]. Henceforth, these results altogether show that the CDC effect is purely of extrinsic origin[22], and that the phononic + electronic contributions ($\varepsilon'_{intrisic}$) account completely for the value of the relaxed radio frequency dielectric constant. No signature of ferroelectric-related behavior (normal, incipient nor relaxor ferroelectricity) was suggested by the polar phonon modes. CDC effect in other compounds, like in the high bandgap semiconductor $CaCu_3Ti_4O_{12}$, has being associated to the presence of intentional or non-intentional charged defects, mainly oxygen vacancies or cationic vacancies, that act as charge carriers with relaxing conductivity (and dielectric constant) behaviors at radio frequencies[47,48]. Analogously, LCMO exhibits a semiconductor character[19,49], with charge ordering and oxygen deficiency as functions of processing conditions[15,50,51].

## 4. Conclusions

Infrared reflectivity spectra of $La_2CoMnO_6$ (LCMO) ceramics obtained from a modified Pechini's method were investigated for the first time. X-ray diffraction and XPS analyses showed that the material crystallized into double perovskite structure, with ordered ($Mn^{4+}/Co^{2+}$) cations, of the $P2_1/n$ space group. Dispersion analysis by a four-parameter semi-quantum model reveals the 13 most intense polar modes that describe the reflectivity spectra. The obtained dispersion parameters show that the low-frequency modes (below 180 cm$^{-1}$) account for *ca.* 50% of the dielectric constant. These modes are directly linked to the motions of ions on B-site, as showed by lattice dynamic calculations, from Wilson's matrices. The small values of the dielectric strengths for most of the oscillators show that the modes are weakly coupled, which leads to low dielectric losses. For the intrinsic dielectric response, we obtained the extrapolated values $\varepsilon'_{intr}$ ~16, and a quality factor $Q_u \times f$ (GHz) ~124 THz, at 10 GHz. The intrinsic dielectric constant matches well with the relaxed radio-frequency values demonstrating that the reported Colossal Dielectric Constant effect in LCMO has a purely extrinsic origin.

## Acknowledgments

The authors are grateful to CNPq, CAPES, Fapemig and Fapema for co-funding this work.

## References

[1] Y. Shimakawa, M. Azuma, and N. Ichikawa, Materials (Basel) **4**, 153 (2011).

[2] M.P. Singh, K.D. Truong, S. Jandl, and P. Fournier, J. Appl. Phys. **107**, 09D917 (2010).

[3] W. Eerenstein, N.D. Mathur, and J.F. Scott, Nature **442**, 759 (2006).

[4] D. Bhadra, G. Masud, and B.K. Chaudhuri, Appl. Phys. Lett. **102**, 072902 (2013).

[5] P. Padhan, P. Leclair, A. Gupta, M.A. Subramanian, and G. Srinivasan, J. Phys. Condens. Matter **21**, 306004 (2009).

[6] M.P. Singh, K.D. Truong, and P. Fournier, Appl. Phys. Lett. **91**, 042504 (2007).

[7] D. Choudhury, P. Mandal, R. Mathieu, A. Hazarika, S. Rajan, A. Sundaresan, U. Waghmare, R. Knut, O. Karis, P. Nordblad, and D. Sarma, Phys. Rev. Lett. **108**, 127201 (2012).

[8] Y.Q. Lin and X.M. Chen, J. Am. Ceram. Soc. **94**, 782 (2011).

[9] N.S. Rogado, J. Li, A. W. Sleight, and M.A. Subramanian, Adv. Mater. **17**, 2225 (2005).

[10] R.N. Mahato, K. Sethupathi, and V. Sankaranarayanan, J. Appl. Phys. **107**, 09D714 (2010).

[11] C.L. Bull, D. Gleeson, and K.S. Knight, J. Phys. Condens. Matter **15**, 4927 (2003).

[12] A.J. Barón-González, C. Frontera, J.L. García-Muñoz, B. Rivas-Murias, and J. Blasco, J. Phys. Condens. Matter **23**, 496003 (2011).

[13] K.D. Truong, M.P. Singh, S. Jandl, and P. Fournier, Phys. Rev. B **80**, 134424 (2009).

[14] K.D. Truong, J. Laverdière, M.P. Singh, S. Jandl, and P. Fournier, Phys. Rev. B **76**, 132413 (2007).

[15] R. Dass and J. Goodenough, Phys. Rev. B **67**, 014401 (2003).

[16] F.N. Sayed, S.N. Achary, O.D. Jayakumar, S.K. Deshpande, P.S.R. Krishna, S. Chatterjee, P. Ayyub, and A.K. Tyagi, J. Mater. Res. **26**, 567 (2011).

[17] Y.Q. Lin and X.M. Chen, J. Am. Ceram. Soc. **94**, 782 (2011).

[18] S. Yáñez-Vilar, M. Sánchez-Andújar, J. Rivas, and M.A. Señarís-Rodríguez, J. Alloys Compd. **485**, 82 (2009).

[19] Y.Q. Lin and X.M. Chen, J. Am. Ceram. Soc. **94**, 782 (2011).

[20] M.A. Subramanian, D. Li, N. Duan, B.A. Reisner, and A.W. Sleight, J. Solid State Chem. **151**, 323 (2000).

[21] C.C. Holmes, T. Vogt, S.M. Shapiro, S. Wakimoto, and A.P. Ramirez, Science (80-. ). **293**, 673 (2001).

[22] C.P.L. Rubinger, R.L. Moreira, G.M. Ribeiro, F.M. Matinaga, S. Autier Laurent, B. Mercey, and R.P.S.M. Lobo, J. Appl. Phys. **110**, 074102 (2011).

[23] A. Venimadhav, D. Chandrasekar, and J.K. Murthy, Appl. Nanosci. **3**, 25 (2013).

[24] H.-L. Liu, H.-C. Hsueh, I.-N. Lin, M.-T. Yang, W.-C. Lee, Y.-C. Chen, C.-T. Chia, and H.-F. Cheng, J. Phys. Condens. Matter **23**, 225901 (2011).

[25] G.S. Babu, V. Subramanian, V.R.K. Murthy, I.-N. Lin, C.-T. Chia, and H.-L. Liu, J. Appl. Phys. **102**, 064906 (2007).

[26] M. Iliev, M. Abrashev, A. Litvinchuk, V. Hadjiev, H. Guo, and A. Gupta, Phys. Rev. B **75**, 104118 (2007).

[27] M.P. Pechini, U.S. patent 3.330.697 (1967).

[28] C.C. Marsh, A. Clark, Philos. Mag. **19**, 449 (1969).

[29] A.C. Larson and R.B. Von Dreele, *General Structure Analysis System (GSAS)* (1994).

[30] B.H. Toby, J. Appl. Crystallogr. **34**, 210 (2001).

[31] M.W. Lufaso and P.M. Woodward, Acta Crystallogr. B. **57**, 725 (2001).

[32] C.L. Bull, D. Gleeson, and K.S. Knight, J. Phys. Condens. Matter **15**, 4927 (2003).

[33] R.X. Silva, H. Reichlova, X. Marti, D.A.B. Barbosa, and M.W. Lufaso, J. Appl. Phys. **114**, 194102 (2013).

[34] R.B. Macedo Filho, A.P. Ayala, and C.W.A. Paschoal, Appl. Phys. Lett. **102**, 192902 (2013).

[35] See supplementary material at [……] for more details about the structure refinement and lattice parameters calculations.

[36] H.Z. Guo, A. Gupta, T.G. Calvarese, and M.A. Subramanian, Appl. Phys. Lett. **89**, 262503 (2006).

[37] J.F. Moulder and J. Chastain, *Handbook of X-Ray Photoelectron Spectroscopy: A Reference Book of Standard Spectra for Identification and Interpretation of XPS Data* (Perkin Elmer Corp., 1992).

[38] M.D. Fontana, G. Metrat, J.L. Servoin, and F. Gervais, J. Phys. C Solid State Phys. **17**, 483 (1984).

[39] M. Furuya, J. Korean Phys. Soc. **32**, S353 (1998).

[40] R. Zurmühlen, J. Petzelt, S. Kamba, V. V. Voitsekhovskii, E. Colla, and N. Setter, J. Appl. Phys. **77**, 5341 (1995).

[41] G. Santosh Babu, V. Subramanian, V.R.K. Murthy, R.L. Moreira, and R.P.S.M. Lobo, J. Appl. Phys. **103**, 084104 (2008).

[42] A. Dias, G. Subodh, M.T. Sebastian, M.M. Lage, and R.L. Moreira, Chem. Mater. **20**, 4347 (2008).

[43] A. Dias, G. Subodh, M.T. Sebastian, and R.L. Moreira, J. Raman Spectrosc. **41**, 702 (2010).

[44] F. Gervais and B. Piriou, Phys.Rev. B Solid State **10**, 1642 (1974).

[45] T. Shimanouchi, M. Tsuboi, and T. Miyazawa, J. Chem. Phys. **35**, 1597 (1961).

[46] L. Piseri and G. Zerbi, J. Mol. Spectrosc. **26**, 254 (1968).

[47] M.C. Ferrarelli, D.C. Sinclair, A.R. West, H.A. Dabkowska, A. Dabkowski, and G.M. Luke, J. Mater. Chem. **19**, 5916 (2009).

[48] G. Deng, N. Xanthopoulos, and P. Muralt, Appl. Phys. Lett. **92**, 172909 (2008).

[49] M. Zhu, Y. Lin, E.W.C. Lo, Q. Wang, and Z. Zhao, Appl. Phys. Lett. 062406 (2012).

[50] T. Kyômen, R. Yamazaki, and M. Itoh, Chem. Mater. **15**, 4798 (2003).

[51] M. Viswanathan, P.S.A. Kumar, V.S. Bhadram, C. Narayana, A.K. Bera, and S.M. Yusuf, J. Physics-Condens. Matter **22**, 1 (2010).